\documentclass[aps,pre,twocolumn,groupedaddress,superscriptaddress,showpacs]{revtex4-1}

\usepackage{amsmath}
\usepackage{amssymb}
\usepackage{graphicx}
\usepackage{float}
\usepackage{color}

\begin{document}

\title{Effect of the number of patches on the growth of networks of patchy colloids on substrates}

  \author{C. S. Dias}
   \email{csdias@fc.ul.pt}
    \affiliation{Departamento de F\'{\i}sica, Faculdade de Ci\^{e}ncias, Universidade de Lisboa, P-1749-016 Lisboa, Portugal, and Centro de F\'isica Te\'orica e Computacional, Universidade de Lisboa, Avenida Professor Gama Pinto 2, P-1649-003 Lisboa, Portugal}

  \author{N. A. M. Ara\'ujo}
   \email{nmaraujo@fc.ul.pt}
   \affiliation{Departamento de F\'{\i}sica, Faculdade de Ci\^{e}ncias, Universidade de Lisboa, P-1749-016 Lisboa, Portugal, and Centro de F\'isica Te\'orica e Computacional, Universidade de Lisboa, Avenida Professor Gama Pinto 2, P-1649-003 Lisboa, Portugal}

  \author{M. M. Telo da Gama}
   \email{margarid@cii.fc.ul.pt}
    \affiliation{Departamento de F\'{\i}sica, Faculdade de Ci\^{e}ncias, Universidade de Lisboa, P-1749-016 Lisboa, Portugal, and Centro de F\'isica Te\'orica e Computacional, Universidade de Lisboa, Avenida Professor Gama Pinto 2, P-1649-003 Lisboa, Portugal}

\begin{abstract}
We investigate numerically the irreversible aggregation of patchy spherical colloids
on a flat substrate. We consider $n$-patch particles and characterize the dependence of the
irreversible aggregation kinetics on $n$. For all values of $n$, considered in 
this study, the growing interface of the aggregate is in the Kardar-Parisi-Zhang universality 
class, although the bulk structure exhibits a rich dependence on $n$.
In particular, the bulk density varies with $n$ and the network is more ordered for particles 
with fewer patches. Preferred orientations of the bonds are also observed for networks of particles with low $n$.

\end{abstract}

\maketitle

\section{Introduction}

% Recent experiments for the aggregation of colloids at the edge of an evaporating drop reveal that the interfacial roughening 
% depends on the colloids shape \cite{Yunker2013,Yunker2013b}. Also drying conditions and particle volume 
% fractions were shown to affect the pinning or depinning of the colloidal 
% contact line \cite{Yang2014}. This increasing number of experiments developed to study 
% nonequilibrium aggregates of colloidal particles raises questions on how the 
% control of valence, given by the use of patchy colloids, can affect 
% the kinetics of these interfaces.

Colloids with attractive patches on their surface have been under the spotlight due to their 
possible role in the development of novel materials with fine tuned mechanical, optical, 
and thermal properties \cite{Pawar2010, Kretzschmar2011, Sacanna2011, Bianchi2011}. 
In recent years, many fabrication techniques have been developed \cite{Yi2013,Wilner2012,Hu2012,Duguet2011,Shum2010,Wang2012,He2012} 
which motivated the development of a wide range of theoretical models 
\cite{Glotzer2004, Ruzicka2011,Bianchi2006,Russo2010,Russo2009,Sciortino2011, Sciortino2007, Bianchi2011,Sokolowski2014,Pizio2014}.
Theoretical \cite{Doppelbauer2010,Marshall2013,Marshall2013a,Tavares2014,Markova2014}
and experimental \cite{Iwashita2013,Iwashita2014} studies have revealed 
a strong dependence of the equilibrium structures of patchy colloids on the valence and the strength of the interactions.
After the quest for the feasibility of the equilibrium model structures the emphasis has shifted
to the kinetics~\cite{Sciortino2009,Corezzi2009,Corezzi2012,Vasilyev2013,Markova2014} 
including the adsorption of films on substrates \cite{Gnan2012,Bernardino2012,Dias2013,Dias2013a,Dias2013b}.

Recent efforts have been focused on the nonequilibrium properties of patchy colloids, 
with emphasis on the structure of the adsorbed films \cite{Dias2013,Dias2013a,Dias2013b} 
and the properties of the growing interface \cite{Dias2014,Dias2014a}. 
In particular, it was shown that, in the limit of irreversible adsorption, growth on substrates is sustained only for 
certain arrangements of the patches on the surface of the colloids, delimited by two absorbing phase 
transitions from thin to thick films \cite{Dias2014}. It was also found that while for isotropic 
sticking colloids and single patch-type colloids the growing interface is in the Kardar-Parisi-Zhang universality class 
\cite{Meakin1998,Vold1963,Vold1959,Dias2014} for colloids with weak and strong bonds the interface is in 
the universality class of Kardar-Parisi-Zhang with quenched disorder, for significantly distinct energies of 
the patch-patch interactions \cite{Dias2014a}.

In this work, we study the influence of the number of patches $n$ 
on the interfacial and bulk properties of the networks on flat substrates.

In the following section we give a description of the model. In Sec.~\ref{sec.results}, we present our results 
and in Sec.~\ref{sec.conc}, we draw some conclusions.

\section{Model}\label{sec.model}

\begin{figure}[t]
   \begin{center}
    \includegraphics[width=0.5\columnwidth]{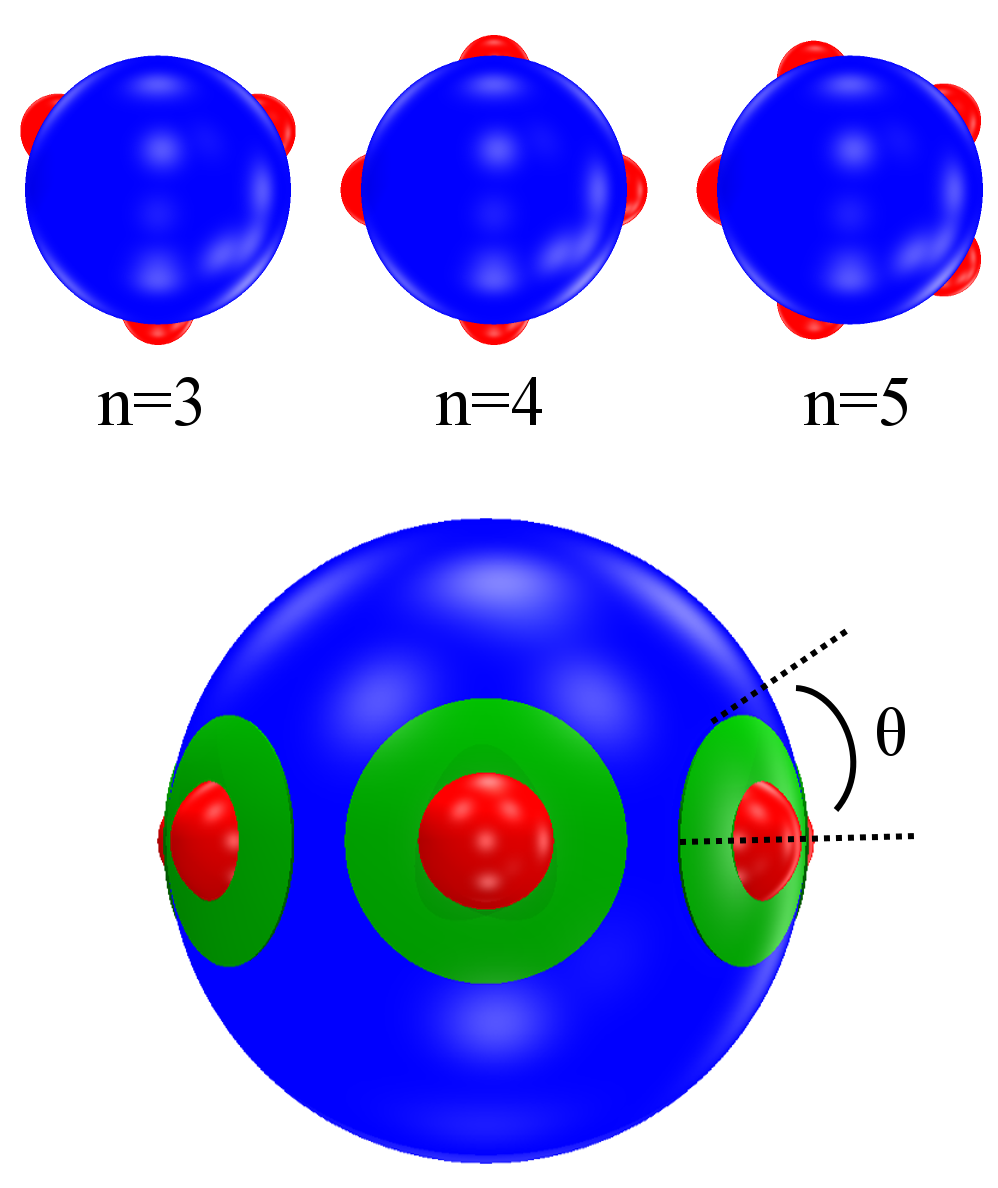} \\
\caption{Bottom: Schematic representation of the patches (red) on the surface of a
colloid (blue) and their interaction range (green). The interaction range is defined by an angle $\theta$ 
with the center of the patch. This angle controls the extension and strength of the patch. 
Top: Patchy colloids with three, four, and five
patches distributed regularly along the equator of the particle.}
  \label{fig.figure1}
   \end{center}
  \end{figure}

We consider spherical three-patch colloids of
unit diameter $\sigma$ with short-range attractive patches 
on their surface. To access large-length and 
long-time scales, we use a stochastic 
model first proposed in Ref.~\cite{Dias2013}.
We describe the patch-patch short-range interaction in a stochastic way and we focus 
on chemical or DNA mediated bonds \cite{Geerts2010,Wang2012}, which 
are highly directional and very strong. This type of bonds may be considered irreversible 
within the timescale of interest \cite{Leunissen2011}, as we consider here.
The high directionality of the interactions is modeled by assuming optimal bonds such 
that the center of two bonded colloids is aligned with their bonding patches.

As in colloidal aggregation at the edge of drops \cite{Yunker2013}, we consider 
advective mass transport only. We denote by $h_\mathrm{max}$ the
maximum height of a colloid in the aggregate and start the simulation from an empty
substrate, i.e. $h_\mathrm{max}=0$. We choose a horizontal position uniformly at
random, at height $h_\mathrm{dep}=h_\mathrm{max}+\sigma$, and simulate the ballistic downward trajectory of the colloid
until it hits either the substrate or another colloid. A colloid-substrate collision results on the adsorption of the 
colloid with a random orientation.

Particle-particle interactions are described, as shown in Fig.~\ref{fig.figure1},
by the interaction range (green), on the surface of the colloid
(blue), around each patch (red). The range is characterized by a single
parameter, namely, its angle $\theta$ with the center of the patch. In a collision with a pre-adsorbed colloid, 
bonding occurs when contact is within overlapping interaction ranges. The position of the incoming particle is then adjusted to
align the patch-patch orientation with that for optimal bonding. Otherwise, bonding fails, i.e., the colloid is discarded
and a new one is released from the top.

\begin{figure}[t]
   \begin{center}
    \includegraphics[width=\columnwidth]{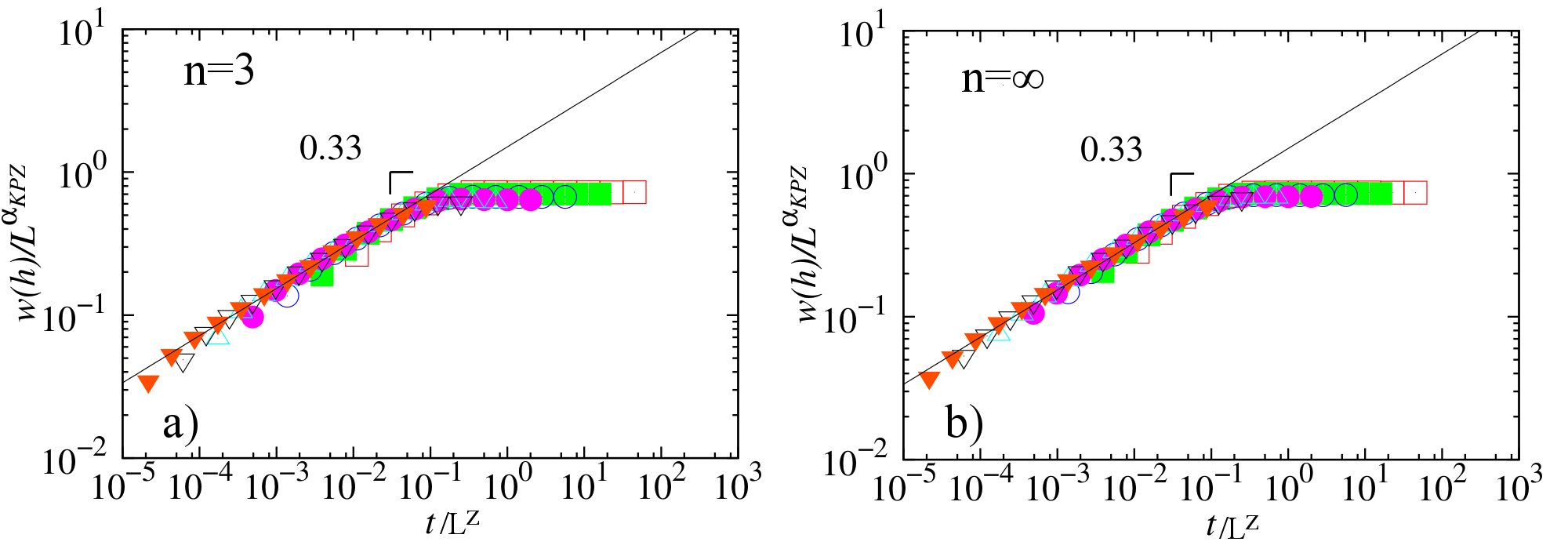} \\
\caption{Data collapse according to the Family-Vicsek scaling relation (\ref{eq:family}) for (a) $n=3$ and 
   (b) $n=\infty$. Simulations were performed on substrates of linear sizes ranging from $L=32$ to $L=2048$ with $2048L$
   adsorbed patchy colloids averaged over $6.4\times 10^5$ samples, for the smaller systems, and $10^4$ samples for the larger ones.
   We considered $\alpha_\mathrm{KPZ}=0.5$ and $z=1.5$ consistent with the Kardar-Parisi-Zhang universality class~\cite{Barabasi1995}.}
  \label{fig.collapse}
   \end{center}
  \end{figure}

\section{Results}\label{sec.results}

% \subsection{Three-patch colloids}
% 
% \begin{figure}[h]
%    \begin{center}
%     \includegraphics[width=\columnwidth]{roughness_angle.pdf} \\
% \caption{Roughness as a function of the opening angle}
%   \label{fig.angle}
%    \end{center}
%   \end{figure}

%\subsection{Increasing number of patches}

In order to investigate the effect of the number of patches $n$, we vary $n$ from 
$n=3$ to $n=\infty$. A schematic representation of the particles is depicted in 
Fig.~\ref{fig.figure1}~(top). For $n=2$ only independent chains grow and there is no interface. 
We view this limit ($n=2$) as the pinned phase. The opposite limit, $n=\infty$, corresponds to the 
aggregation of isotropic sticky colloids \cite{Meakin1998,Vold1963,Vold1959}. We performed detailed 
simulations for systems with substrate sizes ranging from $L=32$ to $L=2048$ and averaged over as 
many as $6.4\times10^5$ samples for the 
smallest systems and $10^4$ samples for the largest ones. Below, we analyze the properties of both the 
interface and the bulk of the adsorbed films. 
We analyze the scaling of the growing interface (Sec.~\ref{sec.interface}), the pair-distribution 
function of the colloids in the network (Sec.~\ref{sec.pair}), 
the density of the bulk (Sec.~\ref{sec.density}), and the orientation of the bonds (Sec.~\ref{sec.angle}).

\subsection{Interfacial scaling}\label{sec.interface}

We start by characterizing the kinetic roughening of the growing interface. We divide the off-lattice system into $N$ 
columns of width $\sigma$, where $N=L/\sigma$. We determine the interfacial height $h_i$, of each column $i$, 
and calculate the interfacial roughness $w$, defined as,

\begin{equation}\label{eq.roughness}
w(t)=\sqrt{\langle\left[h_i(t)-\langle h(t)\rangle\right]^2\rangle}, \\
\end{equation}
where $\langle h(t)\rangle=\sum_i{h_i}/N$ denotes an average over the $N$ columns. The time $t$ is defined as 
the number of adsorbed layers of colloids (equivalent to the average height used in experiments). Initially, 
the roughness increases with time but, due the substrate finite size it saturates eventually at a value 
$w_\mathrm{sat}$ that increases with the system size
\cite{Odor2004,Barabasi1995}. The saturation roughness and saturation time ($t_\mathrm{sat}$) scale with the system size as 
$w_\mathrm{sat}\sim L^{\alpha}$ and $t_\mathrm{sat}\sim L^z$, respectively, where
$\alpha$ is the roughness exponent and $z$ is the dynamical exponent. The short-time behavior of the 
interface roughness is also a power law given by
$w(t)\sim t^\beta$ where $\beta$ is the growth exponent. The interfacial roughness is expected to 
follow the \textit{Family-Vicsek} \cite{Family1985} scaling relation,

  \begin{equation}
     w(L,t)=L^\alpha f\left(\frac{t}{L^z}\right), \\
     \label{eq:family}
  \end{equation} 
  where $f(u)$ is a scaling function. Using the scaling relation and the exponents for different universality classes we can identify 
  the universality class of the growing interface.

\begin{figure}[t]
   \begin{center}
    \includegraphics[width=0.7\columnwidth]{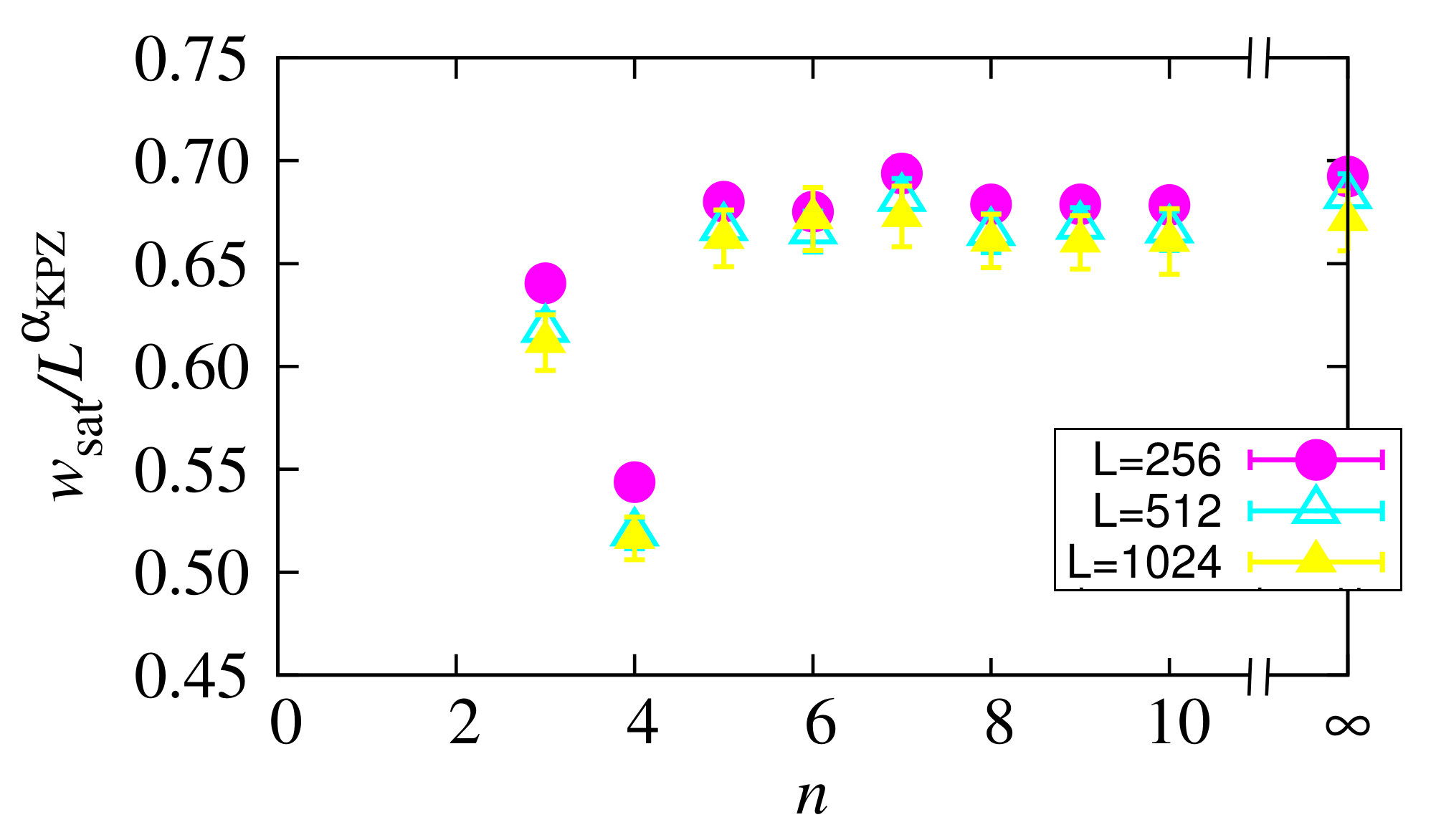} \\
\caption{Data collapse for
the saturation roughness dependence on the number of patches $n=\{3,4,5,6,7,8,9,10,...,\infty\}$,
$w_\mathrm{sat}=L^{\alpha_\mathrm{KPZ}}\mathcal{F}\left[n\right]$,
where $\mathcal{F}$ is a scaling function and $\alpha_\mathrm{KPZ}=0.5$ is
the roughness exponent for the Kardar-Parisi-Zhang universality class.
We considered three different substrate lengths
$L=\{256,512,1024\}$ and results are averages over $\{8\times 10^4,4\times 10^4,2\times 10^4\}$ samples.}
  \label{fig.npatch}
   \end{center}
  \end{figure}

Figure~\ref{fig.collapse} illustrates the data collapse of the 
roughness rescaled using the Family-Vicsek scaling relation (\ref{eq:family}), for 
two values of the number of patches $n=\{3,\infty\}$. The 
data collapse is obtained using the critical exponents of the \textit{Kardar-Parisi-Zhang} 
(KPZ) universality class, namely $\beta=1/3$, $\alpha=1/2$, and $z=3/2$, revealing
that the interface is in this universality class regardless of the anisotropy of the bonding interactions, 
related to the number of patches. The KPZ scaling is corroborated by the scaling of the saturation roughness
for different $n$, illustrated in Fig.~\ref{fig.npatch}, for three different system sizes. Note that data 
collapse is obtained when the roughness is rescaled by $L^{\alpha_\mathrm{KPZ}}$, as expected. 
This result contrasts to that observed for particles which are asymmetric in shape \cite{Yunker2013} 
or bonding energies \cite{Dias2014a}.
We note a remarkable non-monotonic dependence of the roughness on $n$, with a rapid variation for 
low-valence colloids and a minimum at $n=4$, as a result of two distinct mechanisms. While the 
binding probability, increasing with $n$ (maximal for isotropic colloids), increases the interfacial roughening 
there are magic values of $n$, which by favouring local order lead to a strong decrease of the interfacial roughness.

\begin{figure}[t]
   \begin{center}
    \includegraphics[width=0.5\columnwidth]{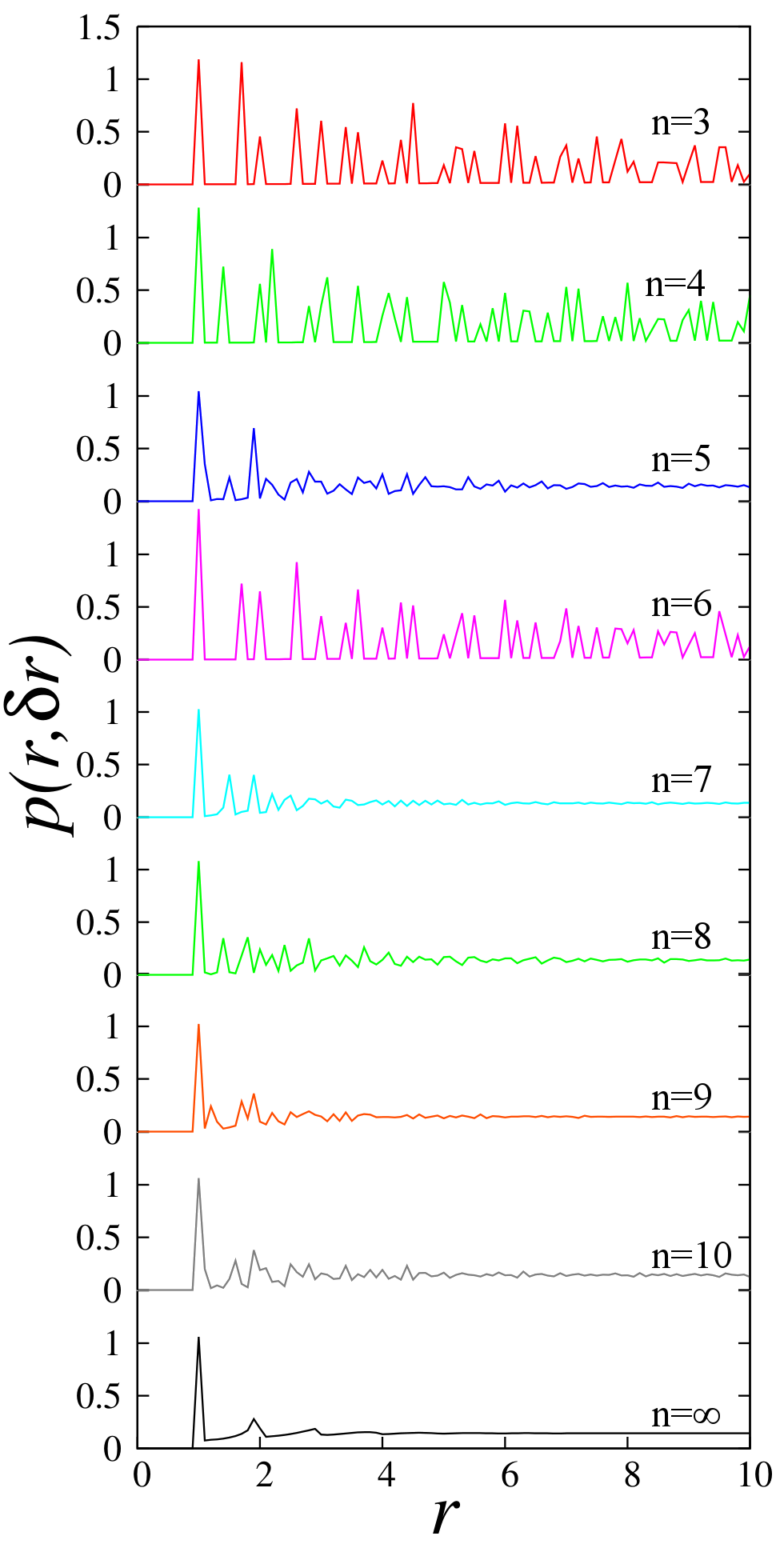} \\
\caption{Pair-distribution function from top to bottom for $n=\{3,4,5,6,7,8,9,10,\infty\}$,
given by Equation~(\ref{eq.pdf}) with $\delta r=0.1$. A system of size $L=512$ and $512L$ 
deposited colloids was considered with the results averaged over $10^4$ samples.}
  \label{fig.PDF}
   \end{center}
  \end{figure}

\subsection{Ordered and disordered structures}\label{sec.pair}
 
A quantitative description of the effect of the number of patches on the local structure of the aggregates is obtained 
from the spatial distribution of the colloids in the network. We used the pair-distribution function 
$p(r,\delta r)$ defined as,
\begin{equation}
p(r,\delta r)=\sum_i^{N-1}\sum_{j+i}^Nf\left(r_{ij}-r,\delta r\right)/Nr,
\label{eq.pdf}
\end{equation}
where $r$ is the distance between each pair $ij$ of colloids and $f\left(r_{ij}-r,\delta r\right)$ is one
if $|r_{ij}-r|<\delta_r$ and zero otherwise. Figure~\ref{fig.PDF} compares the pair-distribution functions 
of the systems investigated ($n=3$ up to $n=\infty$).
For colloids with low valence, the interfacial roughness depends strongly on the local structure of the aggregates.
As revealed by Fig.~\ref{fig.PDF}, the aggregates of colloids with $n=\{3,4,6\}$ exhibit ordered patterns that extend 
over distances much larger than the particle diameter, while for larger values of $n$ only local order is observed.
For colloids with $n>6$, the aggregates are disordered and the structure approaches that of isotropic colloids, 
$n=\infty$, as $n$ increases (see bottom of Fig.~\ref{fig.PDF}). The latter is characterized by a well defined 
peak at $r=1$ (shell of first neighbors) and weak secondary peaks slightly below $r=2$ and $r=3$ 
(shells of second and third neighbors, respectively).

\begin{figure}[t]
   \begin{center}
    \includegraphics[width=0.7\columnwidth]{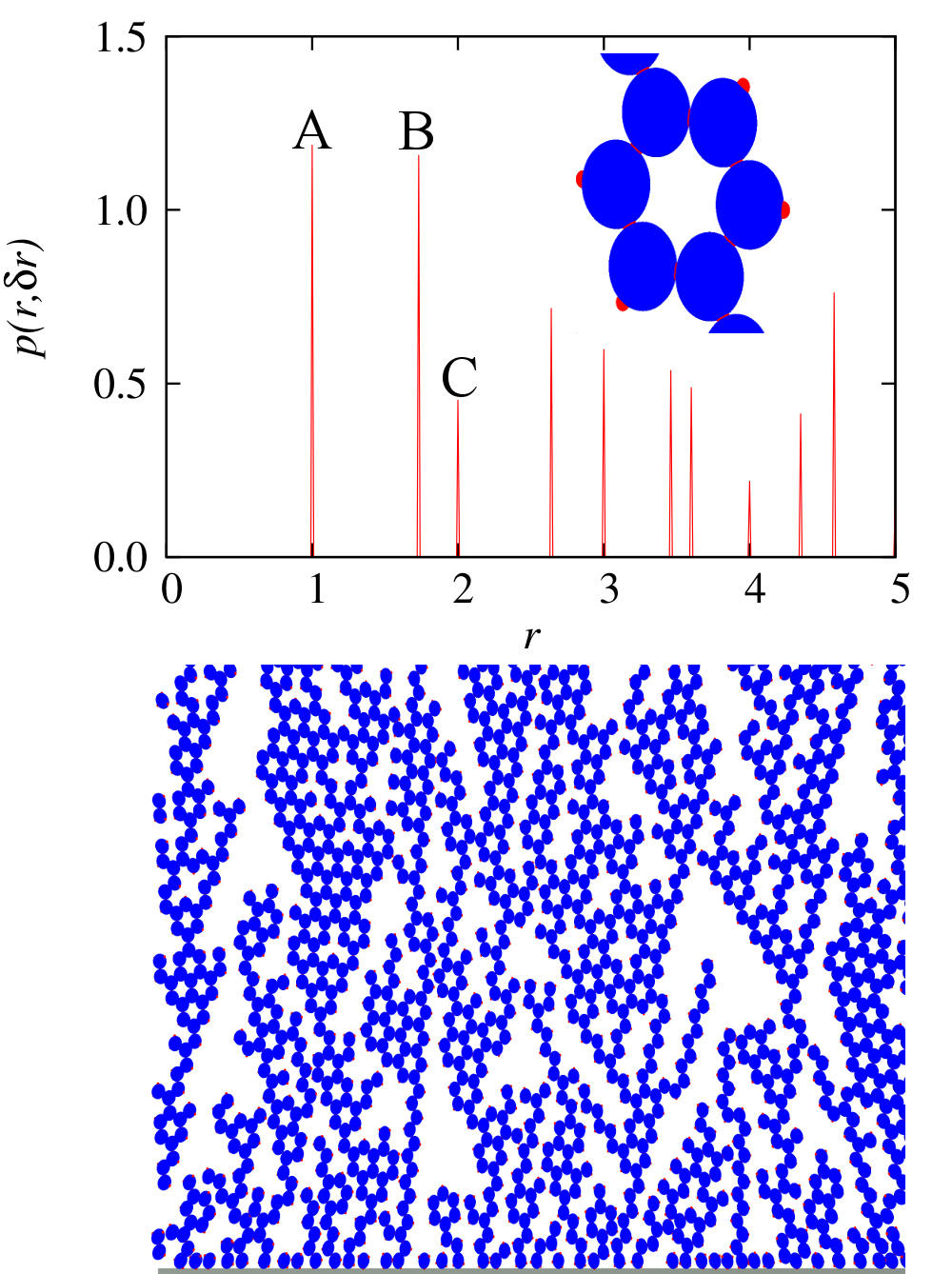} \\
\caption{Top: Pair-distribution function for aggregates of three-patch colloids 
with $\delta r=0.01$.
The peaks A, B, and C are at the distances of the first three neighbors on the honeycomb lattice.
A substrate of length $L=512$ and $512L$ aggregated colloids averaged over $10^4$ samples was used.
Top inset: Zoom of a typical local structure in aggregates of three-patch colloids.
Bottom: Snapshot of a system with substrate length $L=128$ with $10L$ deposited three-patch colloids.}
  \label{fig.PDF_N3}
   \end{center}
  \end{figure}

\begin{figure}[t]
   \begin{center}
    \includegraphics[width=0.7\columnwidth]{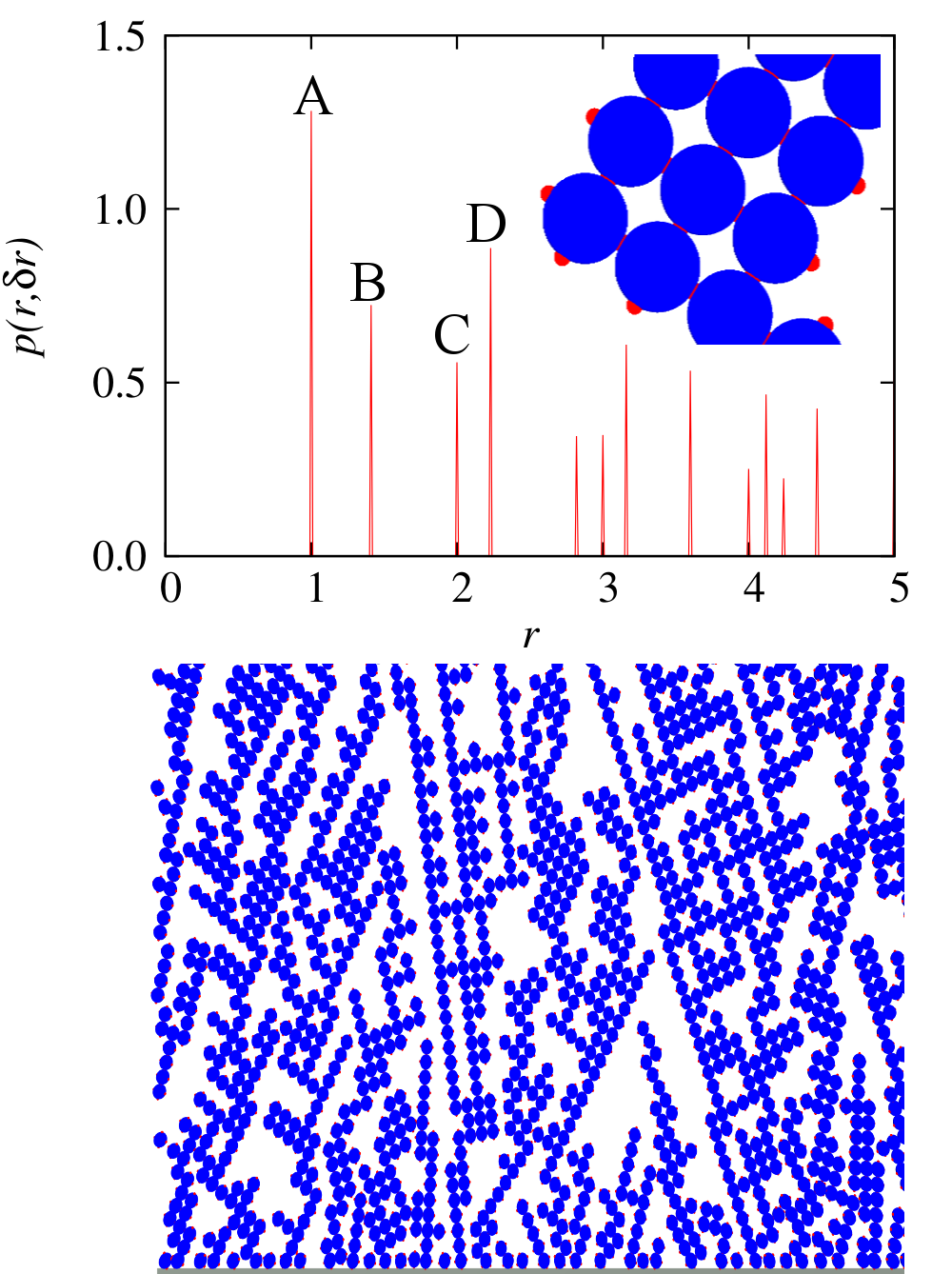} \\
\caption{Top: Pair-distribution function for aggregates of four-patch colloids 
with $\delta r=0.01$.
The peaks A, B, C, and D are at the distances of the first four neighbors on the square lattice.
A substrate of length $L=512$ and $512L$ aggregated colloids averaged over $10^4$ samples was used.
Top inset: Zoom of a typical local structure in aggregates of four-patch colloids.
Bottom: Snapshot of a system of substrate length $L=128$ with $10L$ deposited four-patch colloids.}
  \label{fig.PDF_N4}
   \end{center}
  \end{figure}

\begin{figure}[t]
   \begin{center}
    \includegraphics[width=0.7\columnwidth]{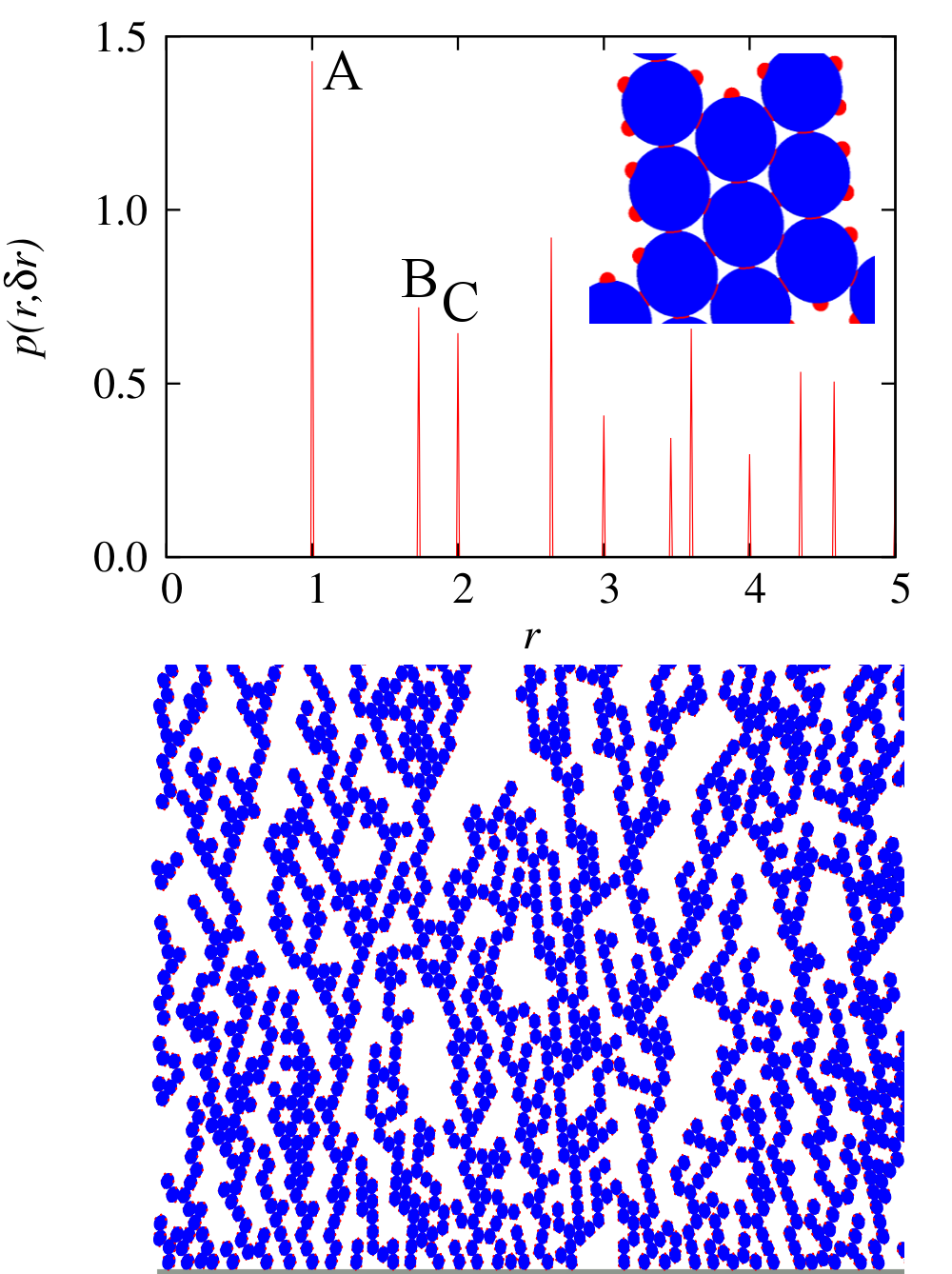} \\
\caption{Top: Pair-distribution function for aggregates of six-patch colloids 
with $\delta r=0.01$.
The peaks A, B, and C are at the distances of the first three neighbors on the triangular lattice.
A substrate of length $L=512$ and $512L$ aggregated colloids averaged over $10^4$ samples was used.
Top inset: Zoom of a typical local structure in aggregates of six-patch colloids.
Bottom: Snapshot of a system of substrate length $L=128$ with $10L$ deposited six-patch colloids.}
  \label{fig.PDF_N6}
   \end{center}
  \end{figure}

To quantify the order of the colloidal aggregates we re-plot the pair distribution functions in  
Figs.~\ref{fig.PDF_N3},~\ref{fig.PDF_N4},~and~\ref{fig.PDF_N6} with smaller binning, namely, $\delta r= 0.01$. 
Figure~\ref{fig.PDF_N3} shows the pair-distribution function for aggregates of three-patch
colloids ($n=3$). The snapshot at the bottom of Fig.~\ref{fig.PDF_N3} reveals that the local structure of 
the nonequilibrium aggregate resembles that of a honeycomb lattice. Furthermore, it is clear that the local 
structure extends over large distances, of the order of the substrate length. The first three peaks of the 
pair-distribution function, A, B, and C, are at the first, second, and third nearest neighbor distances on 
the honeycomb lattice, i.e., $r_A=1$, $r_B=1.73$, and 
$r_C=2$, respectively. The solid-like structure is far from perfect with large irregular voids scattered within ordered domains.

For colloids with $n=4$, local order extending over large distances is also observed (zoom and snapshot in 
Fig.~\ref{fig.PDF_N4}). Since the valence is now four, the local structure is similar to that of the square lattice. 
The first peaks in the pair-distribution function correspond to the first ($r_A=1$), second ($r_B=1.414$), 
third ($r_C=2$), and fourth neighbors ($r_D=2.236$) on the square lattice. The solid-like structure is far 
from perfect with large irregular voids scattered within ordered domains.

Figure~\ref{fig.PDF_N6} shows the results for colloids with $n=6$. We note that, on the large scale, there is no order. 
The local order, however, is still well defined (zoom in top inset of Fig.~\ref{fig.PDF_N6}). 
This is confirmed by the pair-distribution function, plotted in Fig.~\ref{fig.PDF_N6}, with well defined peaks at 
distances corresponding to the first three neighbors on the triangular lattice, $A$ at $r_A=1$, $B$ at $r_B=1.73$, and $C$ at $r_C=2$.

The positions of the first three peaks of the pair-distribution function of six-patch colloids are the same as 
those of three-patch colloids, as observed in honeycomb and triangular lattices. The relative intensity of the peaks, 
however, differs in the two types of aggregates, which is also different from the relative intensity of the peaks in 
the corresponding lattices. The difference in the intensity of the peaks for systems with 
valence three and six results from the increase in the number of neighbors with valence, in line with the 
corresponding lattices.
  
\begin{figure}[t]
   \begin{center}
    \includegraphics[width=0.7\columnwidth]{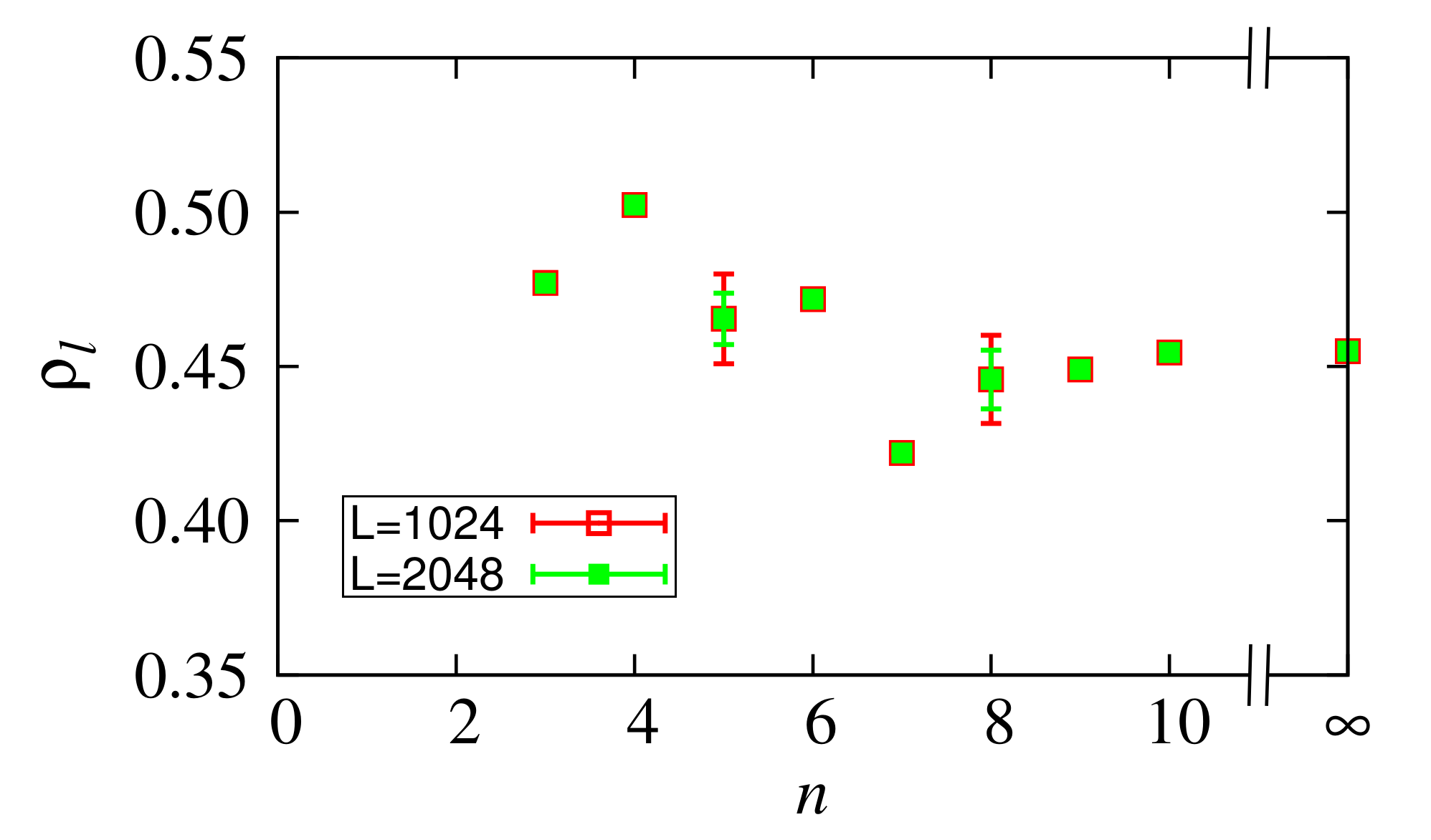} \\
\caption{Density of the film as a function of the number of patches $n$ for systems of 
substrate length $L=\{1024, 2048\}$ and $2048L$ aggregated colloids averaged over $10^5$ samples.}
  \label{fig.density_npatch}
   \end{center}
  \end{figure}

\subsection{Network density}\label{sec.density}
  
One property of practical interest is the network density, which we plot in Fig.~\ref{fig.density_npatch} as a function
of the number of patches $n$. We can identify two regimes: one for systems with low valence where ordered structures
dominate and the other where the aggregates are amorphous. In the ordered regime, the density changes abruptly with $n$. 
The density has a maximum at $n=4$ due to compactness of the local structures when compared with those at $n=3$.
It decreases for $n=5$ as no ordered structures form. For systems with $n=6$ the density increases again as a result of 
local triangular structures. However, the effects of the growth kinetics destroys the global order and thus the 
density is significantly lower than for $n=3$ and $n=4$. Locally the voids may appear larger, as from the
snapshots of both Figs.~\ref{fig.PDF_N3}~and~\ref{fig.PDF_N4}, however ordered structures are less affected by fluctuations
in the structure. On the other hand, for amorphous-like structures the fluctuations in the aggregate structures are higher
and large scale voids are more common, which decreases the global density of the network. We note that the 
density dependence 
on $n$ is opposite to that of the roughness. In fact, denser structures reduce the fluctuations of the interfacial 
height, which in turn decrease the roughness. For amorphous films, however, both the density and the roughness increase 
slightly as $n$ increases, below the limit of isotropic colloids $n=\infty$.

\begin{figure}[t]
   \begin{center}
    \includegraphics[width=0.7\columnwidth]{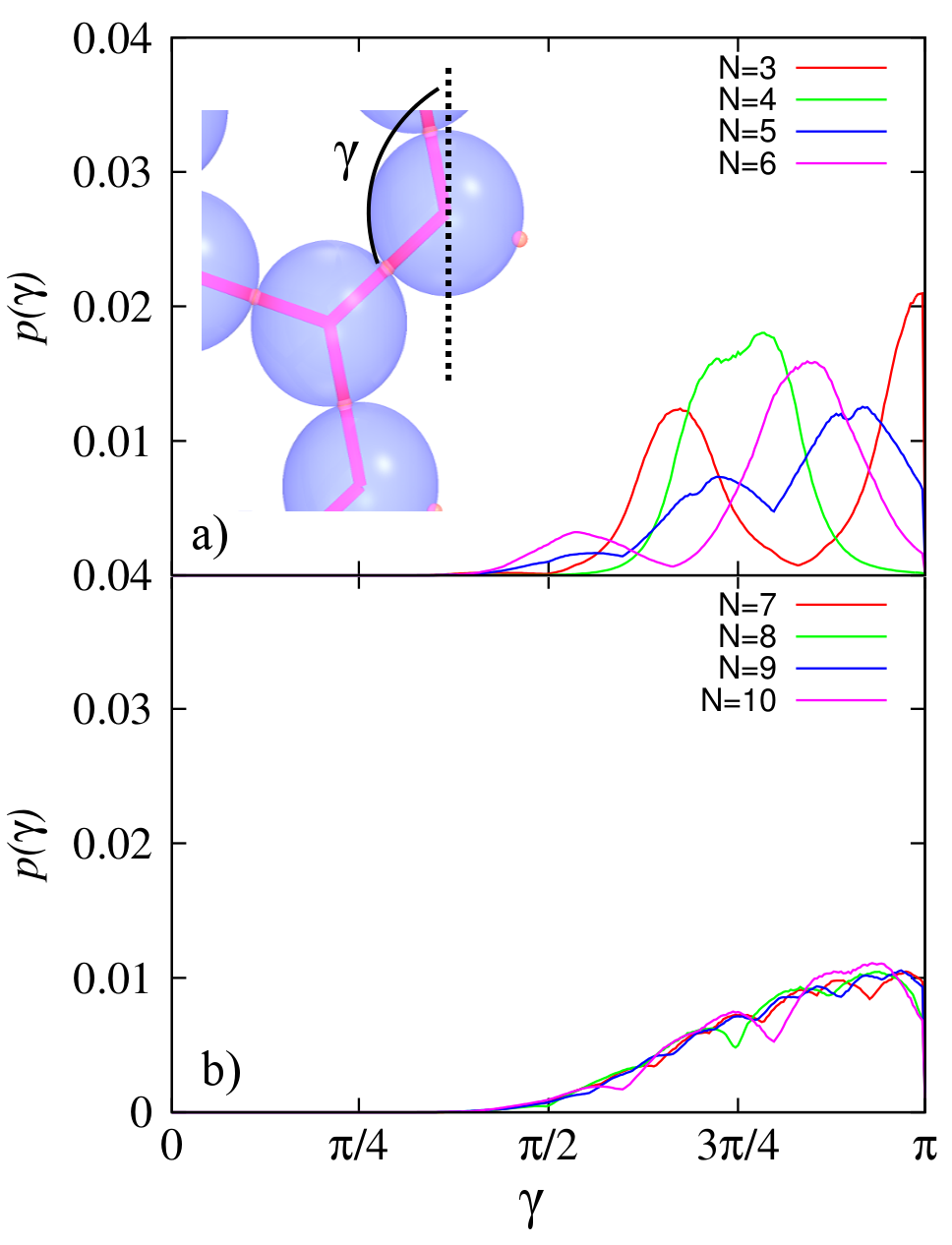} \\
\caption{Probability of the patch-patch bond orientation for a) colloids with $n=\{3,4,5,6\}$ patches and
b) colloids with $n=\{7,8,9,10\}$ patches. The orientation of the bond is measured with respect to a vertical
axis, as shown in the inset of figure a). Simulations were performed for a system of size $L=2048$ with $2048L$ deposited
colloids and averaged over $10^5$ samples.}
  \label{fig.angle_distrib}
   \end{center}
  \end{figure}
  
\subsection{Bond orientation}\label{sec.angle}
  
Finally we address the orientation of the bonds between patches. This analysis gives insight into the kinetics of growth of the 
network of patchy colloids. We consider a vertical $y$ axis on the two dimensional system,
as shown in the inset of Fig.~\ref{fig.angle_distrib}(a) (dashed line) and compute the distribution function of 
the angle between the bond and this direction. The results plotted in Fig.~\ref{fig.angle_distrib}, reveal that 
for systems with low valence, $n<7$ the angular distribution exhibits a sequence of broad but well defined maxima, 
while for systems with high valence the angular distribution is much less structured.

For patchy colloids with highly directional bonds, the orientation of the colloids on the substrate 
is of major importance. The orientation of the colloids in the network depends on the orientational distribution 
of the first layer. Orientation of the patches close to the vertical favors the growth of the network by minimizing 
the blocking effects of colloids in the next layer. Sustained growth is favored by the vertical alignment of 
one patch, since vertically growing aggregates are favored over diagonally and laterally growing ones, 
or maximal exposure of the patches, since aggregates with more exposed patches are more likely to grow.
Geometrical constraints lead to a 
strong dependence on $n$. For $n=3$, Fig.\ref{fig.angle_distrib}(a), the preferred orientation of the 
bonds is $\gamma=2\pi/3$ or $\gamma=\pi$, corresponding to colloids with one patch aligned with the $y$ axis or 
two patches at an angle on either sides of the $y$ axis. For $n=4$, the distribution function has a maximum 
at $\gamma=3\pi/4$, Fig.~\ref{fig.angle_distrib}(a), which corresponds to colloids with two patches at an angle 
on either sides of the $y$ axis. In this 
case the second condition dominates. For colloids with $n=5$ the distribution function,  
Fig.~\ref{fig.angle_distrib}(a), has three maxima indicating that three patches are exposed to newly incoming colloids.
The first maximum is just above $\gamma=\pi/2$ and thus one patch is aligned with the $x$ axis. For systems with $n=6$ the 
alignment is similar to that of three-patch colloids. Up to four patches may be exposed but symmetry reduces the number of maxima 
of the distribution function to two.

Maximal exposure of the free patches determines the preferred orientation(s) of colloids in networks of low valence colloids. 
These are indicated by maxima in Fig.~\ref{fig.angle_distrib}(a) which become less visible for $n>6$, Fig.~\ref{fig.angle_distrib}(b).

%   
% \subsection{Mixtures of patchy colloids}
% 
% \subsubsection{two- and three-patch colloids}
% 
% \begin{figure}[h]
%    \begin{center}
%     \includegraphics[width=\columnwidth]{roughness_dimers.pdf} \\
% \caption{Roughness as a function of the ratio of dimers over trimers}
%   \label{fig.23}
%    \end{center}
%   \end{figure}
% 
% \subsubsection{two- and four-patch colloids}
% 
% \begin{figure}[h]
%    \begin{center}
%     \includegraphics[width=\columnwidth]{roughness_dimers_24.pdf} \\
% \caption{Roughness as a function of the ratio of dimers over tetramers}
%   \label{fig.24}
%    \end{center}
%   \end{figure}

\section{Conclusions}\label{sec.conc}

We studied the aggregation of patchy colloids on substrates and characterized the dependence of the bulk and 
interfacial properties on the number of patches. 
We have shown that regardless of the number of patches the interface is always in the \textit{Kardar-Parisi-Zhang} 
class in contrast to what is observed for selective interaction between patches \cite{Dias2014a}. We found that the 
roughness depends on the number of patches with a minimum at $n=4$, which is similar to what occurs by introducing 
patch-patch correlations on three-patch colloids \cite{Dias2014}.

In addition, we studied the influence of the number of patches on the bulk structure of the film. We have shown that low-valence
colloids aggregate into ordered structures ($n\leq6$) which extend over distances much larger than the diameter of the particles. 
We found
that the local structure of the aggregates and the bulk density depend on the number of patches in a non-monotonic fashion. 
This effect is different to what was reported for mixtures of colloids. In that case, the density increases monotonically 
with the valence, both under equilibrium 
\cite{Bianchi2006,DelasHeras2011} and nonequilibrium \cite{Dias2013b} conditions.

Finally, we measured the orientation of the bonds between patches and found that the valence provides control over the 
directionality of bonds and on the relative growth of the aggregates. This has relevance to
the growth of mixtures of patchy colloids of type $A$ and $B$ with $n$ patches each, where $n$ may be used to tune, for example,
the directionality of dipoles created by the bonds between $A$- and $B$-type colloids. Also the control over the direction 
of conducting circuits created by bonds of patchy colloids could have practical interest.

\begin{acknowledgments}
We acknowledge financial support from the
Portuguese Foundation for Science and Technology (FCT) under Contracts
nos. EXCL/FIS-NAN/0083/2012, PEst-OE/FIS/UI0618/2014, and IF/00255/2013. 
\end{acknowledgments}

\bibliography{colloids_npatch}

\end{document}